\documentclass[preprint,review,12pt]{elsarticle}

\usepackage{amsmath,amssymb,amsfonts}
\usepackage{mathtools}
\usepackage{algorithmic}
\usepackage{bbold}
\usepackage{textcomp}
\usepackage{xcolor}
\usepackage{bm}
\usepackage{array}
\usepackage{booktabs}
\usepackage{tabularx} 
\usepackage{makecell} 
\usepackage[bottom]{footmisc}
\usepackage{xurl}
\usepackage{hyperref}
\usepackage{physics}

\newcommand{\bh}{\ensuremath{\bm{h}}}
\newcommand{\bw}{\ensuremath{\bm{w}}}
\newcommand{\bR}{\ensuremath{\bm{R}}}
\newcommand{\bA}{\ensuremath{\bm{A}}}
\newcommand{\ba}{\ensuremath{\bm{a}}}
\newcommand{\bg}{\ensuremath{\bm{g}}}
\newcommand{\bG}{\ensuremath{\bm{G}}}

\newcommand{\bTheta}{\ensuremath{\bm{\Theta}}}
\newcommand{\btheta}{\ensuremath{\bm{\theta}}}

\newcommand{\be}{\ensuremath{\bm{e}}}

\DeclareMathOperator{\E}{\mathbb{E}}
\DeclareMathOperator{\diag}{diag}

\newlength{\myfigwidth}

\journal{AEU Int. J. of Electronics and Commun.}

\begin{document}
\begin{frontmatter}

\title{Neural Network Based Optimization of Transmit Beamforming and RIS Coefficients Using Channel Covariances in MISO Downlink}

\author[ku]{Khin Thandar Kyaw}
\ead{khinthandar.k@ku.th}

\author[ku]{Wiroonsak Santipach\corref{cor1}}
\ead{wiroonsak.s@ku.ac.th}

\author[km]{Kritsada Mamat}
\ead{kritsada.m@cit.kmutnb.ac.th}

\author[nectec]{Kamol Kaemarungsi}
\ead{kamol.kaemarungsi@nectec.or.th}

\author[titech]{Kazuhiko Fukawa}
\ead{fukawa@radio.ict.e.titech.ac.jp}

\author[cu]{Lunchakorn Wuttisittikulkij}
\ead{Lunchakorn.W@chula.ac.th}

\affiliation[ku]{organization={Department of Electrical Engineering, Faculty of Engineering, Kasetsart University},
             city={Bangkok},
             country={Thailand}}

\affiliation[km]{organization={Department of Electronic Engineering Technology, College of Industrial Technology, King Mongkut’s University of Technology North Bangkok}, 
      city={Bangkok}, 
      country={Thailand}}

\affiliation[nectec]{organization={National Electronics and Computer Technology Center, National Science and Technology Development Agency},
city={Pathumthani}, country={Thailand}}

\affiliation[titech]{organization={Department of Information and Communications Engineering, School of Engineering, Tokyo Institute of Technology}, city={Tokyo}, country={Japan}}

\affiliation[cu]{organization={Department of Electrical Engineering, Faculty of Engineering, Chulalongkorn University},
             city={Bangkok},
             country={Thailand}}

\cortext[cor1]{Corresponding author.}

\begin{abstract}
We propose an unsupervised beamforming neural network (BNN) and a supervised reconfigurable intelligent surface (RIS) convolutional neural network (CNN) to optimize transmit beamforming and RIS coefficients of multi-input single-output (MISO) downlink with RIS assistance.  To avoid frequent beam updates, the proposed BNN and RIS CNN are based on slow-changing channel covariances and are different from most other neural networks that utilize channel instances. Numerical simulations show that the proposed BNN with RIS CNN can achieve much higher sum rates than zeroforcing beamforming with waterfilling power allocation does, especially for systems with higher load, and reduces computation time.
\end{abstract}



\end{frontmatter}



\section{Introduction}

Rapid demand for wireless applications has led to the development of next-generation networks that can accommodate high data rates and wide connectivity. To meet these demands on 6G networks, many promising technologies are being considered, including extremely large-scale multiple-input multiple-output (XL-MIMO)~\cite{wang24, UBIALI2021} and reconfigurable intelligent surfaces (RIS)~\cite{risbenefits21, huang20, jana2023}.  Massive antennas at the base station (BS) can improve spectral and energy efficiency and serve a large number of users simultaneously in the same frequency band~\cite{massiveMIMO13}. However, additional active antennas and expensive radio frequency chains at higher frequencies increase energy consumption and hardware costs, which pose problems for long-term scalability. In contrast, multiple-input single-output (MISO) downlink beamforming in which a BS deploys multiple transmit antennas and each user equipment (UE) has a single receive antenna offers a more efficient alternative to enhance signal quality by aligning transmit beams toward users' direction. Transmit beamforming for UEs can be optimized for various objectives subject to quality of service constraints~\cite{xia20,zhangju20,LAW2017}. The optimized performance depends on the current state of the channel and can be analyzed for certain channel models and limited channel information~\cite{lee16, tcom18}.  

To adapt channels to better suit transmission and improve cell coverage, network operators can install RIS panels in the service area. RIS is a planar surface composed of numerous passive reflective elements connected to a smart controller with each element capable of independently altering the phase of the incident signal it reflects~\cite{risbenefits21,Wu20}. These passive elements, which operate over short ranges, enable RIS to be deployed densely with low energy consumption and reduced costs, without the need for RF chains~\cite{Wu20}. By strategically reflecting or redirecting signals, RIS further improves the signal-to-interference plus noise ratio (SINR) and increases overall network throughput~\cite{Haugn19, qurrat20, nadeem20}. RIS, which does not require a power amplifier, is more energy efficient than a relay that uses amplify and forward or decode and forward schemes~\cite{di20Relay, bjorn20}. 

In this work, we are interested in optimizing both the transmit beamforming at the BS and RIS for the MISO downlink channels~\cite{qurrat20,nadeem20,hu2020statistical,shin2023}. Unlike~\cite{qurrat20,nadeem20}, we assume that only channel covariances of the UE, which vary much more slowly than channel instances, are known. Differing from~\cite{hu2020statistical,shin2023}, we propose applying neural networks to predict optimal transmit beamforming and RIS coefficients that maximize the sum rate over all UEs in downlink transmission.

To optimize BS beamforming and transmit power of all UEs, including UEs with RIS assistance and UEs without RIS assistance, we apply an unsupervised neural network. Training of the network is based on the structure of the transmit beam that is optimal for the channel covariance matrix of any rank. For RIS, we propose the supervised convolutional neural network (CNN) that trains on the solutions obtained from channel covariances using the alternating direction method of multipliers (ADMM)~\cite{boyd11}. The proposed RIS CNN performs close to ADMM with much less computational complexity.  With differing loads and all SNR ranges shown by numerical results, the sum rate obtained by our proposed schemes is higher than that by zeroforcing beamforming with waterfilling power allocation, which is near optimum for systems with low number of UEs or large SNR. The computation time of the proposed scheme is shown to be shorter than that of the zeroforcing method and much shorter than that of ADMM.

\section{Transmission Model and Problem Formulation}
\label{channelModel}

\begin{figure}[htb]
\centering
\includegraphics[width=\myfigwidth]{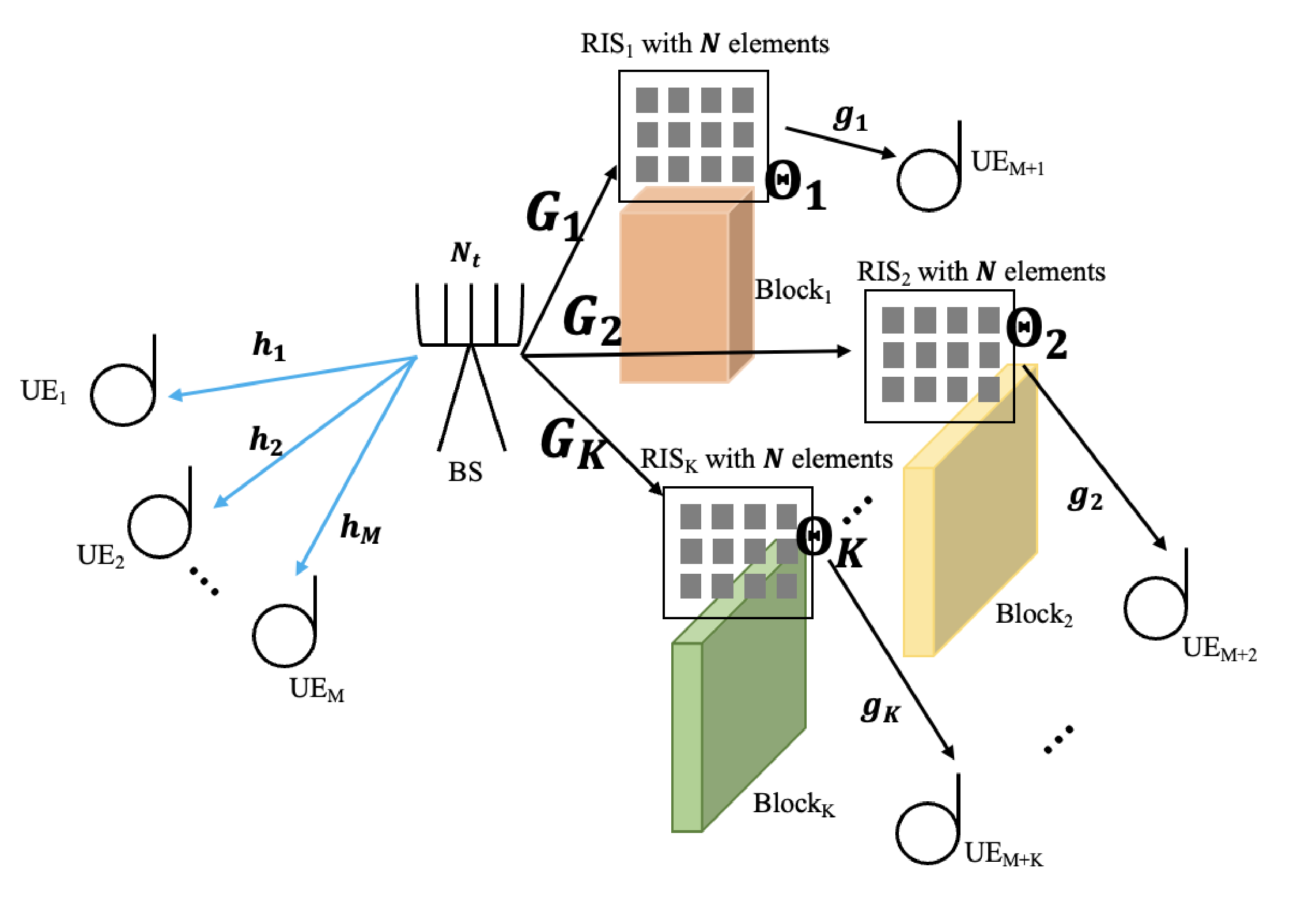}
\caption{A MISO downlink with a BS with $N_t$ antennas, $M$ direct-path UEs, and $K$ UEs with RIS assistance. Each UE has a single receive antenna.}
\label{fig1}
\end{figure}

We consider a downlink transmission in which a BS with $N_t$ antennas transmits to $M+K$ single antenna UEs where $M$ and $K$ denote the number of direct-path UEs and the number of UEs blocked from the BS. Each of the $K$ UEs is assisted by an RIS with $N$ reflecting elements. The MISO downlink considered is illustrated by Fig.~\ref{fig1}. We note that there are at least $K$ RIS panels that assist $K$ UEs that are blocked from the BS. 

With the assumption of flat fading, propagation between each BS transmit antenna to a direct-path UE can be modeled by a complex gain. The BS applies antenna beamforming to transmit message symbols to UEs. The symbol received at UE $i$ is
\begin{equation}
    y_i = \bh_i^{\dagger} \left( \sum_{l=1}^{M+K} \bw_l x_l \right)  + n_i
\label{eq1}
\end{equation}
where $\bh_i$ is an $N_t \times 1$ channel vector whose entries are complex channel gains from all transmit antennas of the BS to the receive antenna of UE $i$, $\bw_l$ is an $N_t \times 1$ beamforming vector for UE $l$, $x_l$ is the message symbol for UE $l$ with zero mean and unit variance, $n_i$ is additive white Gaussian noise symbol with zero mean and $\cdot^{\dagger}$ denotes the Hermitian transpose. Elements of $\bw_l, \forall l$, are complex beamforming coefficients that need to be optimized. With a large frequency reuse factor, interference from co-channel cells is negligible. Therefore, an instantaneous SINR for UE $i$ is given by
\begin{equation}
  \gamma_i = \frac{|\bh_i^{\dagger} \bw_i|^2}{\sum_{\substack{l =1\\ l \ne i}}^{M+K}  |\bh^{\dagger}_i \bw_l|^2 + 1} \label{eq2}
\end{equation}
where we scale the transmit power for UE $i$ dictated by $\|\bw_i\|$ such that the noise power is normalized. 

Instead of instantaneous SINR in~\eqref{eq2}, which varies with channel instance, we consider the expected SINR $\E[\gamma_i]$, which can be approximated by
\begin{equation}
    \tilde{\gamma}_i = \frac{\E [|\bh_i^{\dagger} \bw_i|^2]}{ \E [ \sum_{l \ne i}  |\bh^{\dagger}_i \bw_l|^2] + 1} = \frac{ \bw_i^{\dagger}\bR_{h_i} \bw_i}{\sum_{l \ne i}  \bw_{l}^{\dagger} \bR_{h_i} \bw_{l} + 1} \label{eq_appxsinr}
\end{equation} 
where $\E[\cdot]$ is an expectation operator and $\bR_{h_i} = \E [ \bh_i \bh_i^\dag]$ is the $N_t \times N_t$ channel covariance of UE $i$, which changes much slower than the channel instance. The covariance matrix $\bR_{h_i}$ is positive definite with $\rank(\bR_{h_i}) = L_i > 0$ where $L_i$ corresponds to the number of signal paths between BS and UE $i$. 

For UEs with RIS assistance, we adopt the composite channel model from~\cite{nadeem20}.  Suppose $\bG_k \in \mathbb{C}^{N_t \times N} $, $\btheta_k \in \mathbb{C}^{N \times 1}$, and $\bg_k \in \mathbb{C}^{N \times 1}$  denote a channel matrix between BS and the RIS panel that assists UE $k$, a vector of $N$ reflecting coefficients of that RIS panel, and a channel vector between RIS and UE $k$, respectively.  Since the RIS panels considered in this work are passive, the reflected signals will not be directly amplified. Hence, the magnitude of all the RIS coefficients is constant. The $n$th element of RIS $k$ is $[\btheta_k]_n = \frac{1}{\sqrt{N}} \mathrm{e}^{j\theta_{k,n}}$ with an equal magnitude of $\frac{1}{\sqrt{N}}$ where $\theta_{k,n}$ denotes the phase of the $n$th reflecting element of RIS panel $k$. Note that $\btheta_k$ has unit norm for all $k$. Although the magnitudes of all RIS elements are equal, their phases will be adapted to increase performance at UEs. The composite channel of RIS-assist UE $k$ is given by 
\begin{equation}
  \bh_k = \bG_k \bTheta_k \bg_k \label{eq5}
\end{equation}
where the $N \times N$ diagonal matrix $\bTheta_k = \diag\{\btheta_k\}$ and $\diag\{\bm{a}\}$ denotes the operator that diagonalizes the entries of vector $\bm{a}$ .  The channel matrix of the BS-RIS link $\bG_k$ is less dynamic than that of the RIS-UE link $\bg_k$ and hence, is approximately constant for the transmission duration. Assuming independence between $\bG_k$ and $\bg_k$, the channel covariance of the $k$th RIS-assist UE is
\begin{equation}
\bR_{h_k} = \E[\bG_k \bTheta_k \bg_k \bg_k^{\dagger}\bTheta_k^{\dagger} \bG_k^{\dagger}] = \bG_k \bTheta_k \underbrace{\E[\bg_k \bg_k^{\dagger}]}_{\bR_{g_k}} \bTheta_k^{\dagger} \bG_k^{\dagger}  \label{eq6}
\end{equation}
where $\bR_{g_k}$ is the $N \times N$ covariance matrix of the RIS-UE link for the $k$th RIS-assist UE.

We would like to maximize a sum-achievable rate of all UEs over BS beamforming vectors and RIS coefficients under a total-transmit power constraint and equal magnitude of all RIS elements. For tractability, we approximate the sum rate with the approximate expected SINR~\eqref{eq_appxsinr}. The problem can be stated as follows.
\begin{subequations}
  \begin{alignat}{2}
    \max_{\substack{\bw_1, \bw_2, \cdots, \bw_{M+K}\\
    \btheta_{M+1}, \btheta_{M+2}, \cdots, \btheta_{M+K}}} & \quad && \sum_{i=1}^{M+K} \log_2(1 + \tilde{\gamma}_i)\\
    \text{subject to} &&&
    \sum_{i=1}^{M+K} \|\bw_i\|^2 \le P_{T},\\
    &&& | [\btheta_k]_n | = \frac{1}{\sqrt{N}}, \forall n, k = M+1, M+2, \ldots, M+K, \label{eq:constmod}
  \end{alignat}
  \label{eq:opt_w}%
\end{subequations}
where $P_T$ is the total transmit power from the BS. A similar problem was studied by~\cite{xia20} under the assumption that all channel realizations are known.  In this work, we assume only that channel covariances are known to the BS. The joint optimization problem~\eqref{eq:opt_w} is non-convex due to constant modulus constraints~\eqref{eq:constmod}.

In this work, we propose an alternative optimization solution in which the RIS phases are optimized first, and the BS beams are then solved with the optimized RIS coefficients. The solutions are suboptimal but can outperform zeroforcing beamforming. To reduce computational complexity, we applied neural network to predict BS beams and RIS coefficients. The proposed schemes are described in the next two sections. 

\section{Optimizing RIS Coefficients With Convolution Neural Network}

We first optimize the RIS coefficients by maximizing the expected power of the composite channels for UEs with RIS assistance. As a result, the received signal power for these UEs will increase, and so will the expected SINR~\eqref{eq_appxsinr} and the sum rate~\eqref{eq:opt_w}. An expression of the composite channel for UE $k$ in~\eqref{eq5} can be rearranged as follows:
\begin{equation}
    \bh_k = \bG_k \diag\{\bg_k\} \btheta_k.
\end{equation}
Therefore, the expected channel power of RIS-assist UE $k$ is given by
\begin{align}
    \E[\| \bh_k \|^2 ] &= \btheta_k^\dag \E[ \diag\{\bg_k^\ast \} \bG_k^\dag \bG_k \diag\{\bg_k\}] \btheta_k \label{eq:bEG} \\
    &=\btheta_k^\dag (\bR_{g_k}^T \odot [\bG_k^\dag \bG_k])\btheta_k \label{eq:kbG}
\end{align}
where \eqref{eq:bEG} is transformed to~\eqref{eq:kbG} by an element-wise matrix multiplication denoted by $\odot$, and $\cdot^\ast$ and $\cdot^T$ denote complex conjugate and matrix transpose, respectively.
 
Maximizing the expected channel power for UE $k$ given by~\eqref{eq:kbG} can be solved by the following quadratic program subject to constant modulus constraints 
\begin{subequations}
\begin{alignat}{2}
    \max_{\btheta_k \in \mathbb{C}^N} \quad \btheta_k^{\dag} (\bR_{g_k}^T \odot [\bG_k^{\dag} \bG_k]) \btheta_k \label{eq:opt_RIS_max}\\
    \text{subject to} \quad
    | [\btheta_k]_n | = \frac{1}{\sqrt{N}}, \ n = 1, 2, \dots, N . \label{eq:amp_const}
    \end{alignat}
  \label{eq:opt_RIS}%
\end{subequations}  
Similar to the overall problem~\eqref{eq:opt_w}, this RIS optimization is non-convex due to due the constant modulus. The problem can be solved using ADMM~\cite{boyd11}. We apply the implementation of ADMM presented in~\cite[Algorithm 3]{li20} to solve~\eqref{eq:opt_RIS}. ADMM is iterative and consists of $\mathcal{O}(tN)$ additions and $\mathcal{O}(tN^2 + N^3)$ multiplications, where $t$ is the number of iterations~\cite{li20}.

ADMM typically requires multiple iterations to converge, making it computationally expensive, particularly in real-time or large-scale scenarios. It poses challenges for practical deployment, where rapid adjustments are necessary. In contrast, neural networks offer a viable solution to this problem. Once trained, neural networks can provide near-instantaneous predictions of RIS coefficients. However, traditional artificial neural networks (ANN) are not well suited to predict the optimal complex-valued RIS coefficients due to the limitations in handling complex numbers directly. To overcome this challenge, we propose to apply the convolutional neural network to predict the RIS coefficients (RIS CNN). Creating training samples with ADMM is computationally intensive, but can be performed offline. After the optimal RIS-coefficient vector $\btheta_k$ is obtained for each set of $\bG_k$ and $\bR_{g_k}$, the phase of each vector element is given by
\begin{equation}
    \theta_{k,n} = 
    \begin{cases}
        \arctan(\frac{\Im{[\theta_{k}]_n}}{\Re{[\theta_k]_n}}), &\Re{[\theta_k]_n} > 0 \\
        \arctan(\frac{\Im{[\theta_k]_n}}{\Re{[\theta_k]_n}}) + \pi, &\Re{[\theta_k]_n} < 0 \text{ and } \Im{[\theta_k]_n} \ge 0 \\
        \arctan(\frac{\Im{[\theta_k]_n}}{\Re{[\theta_k]_n}}) - \pi, &\Re{[\theta_k]_n} < 0  \text{ and } \Im{[\theta_k]_n} < 0 \\
        \frac{\pi}{2}, \quad 
        &\Re{[\theta_k]_n} = 0 \text{ and } \Im{[\theta_k]_n} > 0 \\
        -\frac{\pi}{2}, &\Re{[\theta_k]_n} = 0 \text{ and } \Im{[\theta_k]_n} < 0
    \end{cases}
    \label{eq:phi}
\end{equation}
where $-\pi < \theta_{k, n} \le \pi, \forall n$. 

The RIS CNN is trained to predict $\hat{\theta}_{k, n}$ using the input features from $\bG_k$, $\bR_{g_k}$, and $\theta_{k,n}$. The complex $\bG_k$, and $\bR_{g_k}$ are decomposed into real and imaginary parts and are represented by two-dimensional arrays of real numbers. The arrays of $\bG_k$, $\bR_{g_k}$ are fed into the first and second input layers, respectively. The dimension of each layer is shown in Table~\ref{tab3}.  To utilize these inputs, the decomposed $\bG_k$ and $\bR_{g_k}$ are reshaped by merging their second and third dimensions, followed by the fourth and fifth dimensions.  The sample size, which corresponds to the first dimension, is managed automatically in batches during training, which is not explicitly described in Table~\ref{tab3}. To extract meaningful features from the real and imaginary parts of these input matrices, two-dimensional convolutional layers are employed. Both Conv2D layers $1$ and $2$ utilize $64$ filters with a $3 \times 3$ kernel size. These layers allow the network to learn localized features effectively. Then, MaxPooling2D layers are applied to downsample the spatial dimensions by half to reduce computational complexity and still retain the most critical features. The outputs from MaxPooling2D layers are concatenated with flattened $\theta_{k,n}$ and processed through alternating batch normalization and dense layers. Based on our empirical results, rectified linear unit (ReLU) activation functions are used in the dense layers. The final dense layer outputs predicted phases $\hat{\theta}_{k, n}$ for $n = 1, 2, \ldots, N$ with a linear activation function.

\begin{table}
\caption{Output dimensions and activation functions of all layers of the proposed RIS CNN.}
\label{tab3}
\begin{center}
\newcolumntype{C}{>{\centering\arraybackslash}X} 
\begin{tabularx}{0.9\columnwidth}{|c|C|c|} 
\hline
\textbf{Layer number} & \textbf{Output dimension} & \textbf{Activation function} \\ \hline
Input layer 1 & $[K, 2, N_t, N]$ & $\diagdown$  \\ 
\hline
Input layer 2 & $[K, 2, N, N]$ & $\diagdown$ \\ 
\hline
Reshape layer 1 & $[K \times 2, N_t, N]$  & $\diagdown$ \\
\hline
Reshape layer 2 & $[K \times 2, N, N]$  & $\diagdown$ \\
\hline
Conv2D layer 1 & $[K \times 2, N_t, 64]$ & $\diagdown$  \\
\hline
Conv2D layer 1 & $[K \times 2, N, 64]$ & $\diagdown$  \\
\hline
MaxPooling2D layer 1 & $[\frac{K \times 2}{2}, \frac{N_t}{2}, 64]$ & $\diagdown$  \\
\hline
MaxPooling2D layer 2 & $[\frac{K \times 2}{2}, \frac{N}{2}, 64]$ & $\diagdown$  \\
\hline
Input layer 3 & $[K, N, 1]$ & $\diagdown$ \\ 
\hline
Concatenate layer & $[(K \times \frac{N_t}{2} \times 64) + (K \times \frac{N}{2} \times 64) + (K \times N), 1]$ & $\diagdown$ \\ \hline
Dense layer 1 & $[512, 1]$ & ReLU \\ \hline 
Dense layer 2 & $[256, 1]$ & ReLU \\ \hline
Dense layer 3 & $[128, 1]$ & ReLU \\ \hline
Dense layer 4 & $[K \times N]$ & linear \\ \hline
Lambda layer 1 & $[K, N, 1]$ & $\diagdown$ \\ \hline
Lambda layer 2 & \([1]\) & $\diagdown$ \\ \hline
\end{tabularx}
\end{center}
\end{table}

To train the network, we minimize the mean absolute error between the optimal phases and the predicted phases. The loss function of this RIS CNN is given by
\begin{equation}
    \text{Loss} = \frac{1}{N} \sum_{n=1}^N |\theta_{k, n} - \hat{\theta}_{k, n}| .
    \label{eq:loss_RIS_CNN}
\end{equation} 
The RIS coefficients predicted by RIS CNN are given by
\begin{equation}
    \hat{\btheta}_k = \frac{1}{\sqrt{N}} \begin{bmatrix} \mathrm{e}^{j\hat{\theta}_{k, 1}} & \mathrm{e}^{j\hat{\theta}_{k, 2}} & \cdots & \mathrm{e}^{j\hat{\theta}_{k, N}} \end{bmatrix}^T .
    \label{eq:final_RIS_CNN}
\end{equation}
For the model, the dimension of the array containing $\hat{\theta}_{k, n}$ is reshaped in the first lambda layer and the loss function~\eqref{eq:loss_RIS_CNN} is calculated in the second lambda layer. The output dimensions and activation functions of all layers are shown in Table~\ref{tab3}.

For the computation complexity of this network, we can examine the number of floating-point operations per second (FLOPs). For a convolutional layer, the number of FLOPs can be determined from the number of input and output channels, feature height and width, and kernel size of the layer~\cite{flops}. With TensorFlow's default setting, the number of FLOPs of the first and second Conv2D layers in our model are $1152 N K N_t$ and $1152 N^2 K$, respectively.  The complexity of these layers increases with the numbers of RIS-assist UEs, BS antennas, and RIS elements in a panel. MaxPooling layers do not involve multiplications and the associated complexity is negligible. For each dense layer, the number of FLOPs is given by $(2N_I - 1)N_O$ or increases at the rate of $\mathcal{O}(N_I N_O)$, where $N_I$ and $N_O$ denote the input and output dimensions, respectively~\cite{Mol2026,TianLin}. 

To solve problem~\eqref{eq:opt_RIS} with $N_t = 16$ BS antennas, $N = 30$ reflecting elements per RIS panel, and $K = 5$ UEs, the number of FLOPs for the first and second Conv2D layers is approximately $2.76$ million and $5.18$ million, respectively. The number of FLOPs for dense layers are tabulated in Table~\ref{tab4}. The total number of FLOPs for this RIS CNN setup is approximately 16.01 million. 

\begin{table}
\caption{The number of FLOPs for each dense layer for RIS CNN with $N_t = 16$, $N =30$, and $K = 5$.}
\label{tab4}
\begin{center}
\newcolumntype{C}{>{\centering\arraybackslash}X} 
\begin{tabularx}{0.9\columnwidth}{|c|C|c|c|} 
\hline
\textbf{Layer number} & \textbf{Input size} & \textbf{Output size} & \textbf{Number of FLOPs} \\ \hline
Dense layer 1 & $7,510$ & $512$ & $7, 690, 240$ \\ \hline
Dense layer 2 & $512$ & $256$ & $262,144$ \\ \hline
Dense layer 3 & $256$ & $128$ & $65,536$ \\ \hline
Dense layer 4 & $128$ & $150$ & $38,400$ \\ \hline
\end{tabularx}
\end{center}
\end{table}

\section{Optimizing BS Beamforming}
\label{optimization}

After the RIS coefficients are obtained by the RIS CNN and applied, the following subproblem is solved for the BS beamforming vectors:
\begin{subequations}
  \begin{alignat}{2}
    \max_{\bw_1, \bw_2, \cdots, \bw_{M+K}} & \quad && \sum_{i=1}^{M+K} \log_2(1 + \tilde{\gamma}_i)\\
    \text{subject to} &&&
    \sum_{i=1}^{M+K} \|\bw_i\|^2 \le P_{T}.
  \end{alignat}
  \label{eq:subopt_w}%
\end{subequations}
A similar problem was studied by~\cite{xia20} under the assumption that all channel realizations are known.  In this work, we assume only that channel covariances are known to the BS. Similarly to~\cite{bjornson13}, we can show from the dual problem that the optimal solutions to~\eqref{eq:subopt_w} must satisfy the following general eigenvalue-eigenvector equations for $i = 1,2, \ldots, M+K$,
\begin{equation}
    \bR_{h_i} \bw_i = \zeta_i^{\mathrm{\max}} \left( \bm{I} + \sum_{l=1}^{M+K} \beta_l \bR_{h_l} \right) \bw_i
    \label{eq_eig}
\end{equation}
where $\bm{I}$ is the identity matrix of order $N_t$, $\beta_l$'s are the positive parameters with constraint $\sum_{l=1}^{M+K} \beta_l \le P_T$. From~\eqref{eq_eig}, the optimal beam $\bw_i$ is the generalized eigenvector of $(\bR_{h_i}, \bm{I} + \sum_{l=1}^{M+K} \beta_l \bR_{h_l})$ corresponding to the maximum generalized eigenvalue.

If the channel for UE $i$ is sparse with rank-one covariance matrix $\bR_{h_i}$
\begin{equation}
   \bR_{h_i} = \eta_i^{\mathrm{max}} \be_i^{\mathrm{\max}} {\be_i^{\mathrm{\max}}}^\dag 
   \label{eq_rank1}
\end{equation}
where $\be_i^{\mathrm{\max}}$ is the eigenvector of $\bR_{h_i}$ corresponding to the maximum eigenvalue $\eta_i^{\mathrm{max}}$. Substituting~\eqref{eq_rank1} into~\eqref{eq_eig} to obtain
\begin{equation}
    \bw_i = \underbrace{\frac{\eta_i^{\mathrm{max}}{\be_i^{\mathrm{\max}}}^\dag \bw_i}{\zeta_i^{\mathrm{\max}}}}_{\text{Scalar}} \left( \bm{I} + \sum_{l=1}^{M+K} \beta_l \bR_{h_l} \right)^{-1} \be_i^{\mathrm{\max}}.
\end{equation}
Therefore, the optimal beam must have the following form:
\begin{equation}
    \bw_i = \frac{\sqrt{p_i}(\bm{I} + \sum_{l=1}^{M+K} \beta_l \bR_{h_l})^{-1} \be_i^{\max}}{\|(\bm{I} + \sum_{l=1}^{M+K} \beta_l \bR_{h_l})^{-1} \be_i^{\max} \|}
    \label{eq11}
\end{equation}
where $p_i = \|\bw_i\|^2$ is the transmit power allocated to UE $i$. Although~\eqref{eq11} is only optimal when $\rank (\bR_{h_i}) = 1$, it also approximates well when the rank of $\bR_{h_i}$ is small. With~\eqref{eq11}, we can formulate an equivalent problem to~\eqref{eq:opt_w} as follows:
\begin{subequations}
  \begin{alignat}{2}
    \max_{\substack{\beta_1, \beta_2, \cdots, \beta_{M+K}\\
    p_1,p_2, \cdots, p_{M+K}}} & \quad && \sum_{i=1}^{M+K} \log_2(1 + \tilde{\gamma}_i)\\
    \text{subject to} &&&
    \sum_{l=1}^{M+K} p_l \le P_{T},\label{const_p}\\
    &&& \sum_{l=1}^{M+K} \beta_l \le P_{T}, \label{const_b}\\
    &&& \beta_l, \  p_l > 0, \ l = 1,2, \ldots, M+K . \label{const_pos}
  \end{alignat}
  \label{eq:opt_bp}%
\end{subequations}
The number of optimizing parameters for~\eqref{eq:opt_bp} is only $2(M+K)$ while that for~\eqref{eq:opt_w} is $N_t(M+K)$. The difference is large because $N_t$ is typically much greater than $2$. Therefore, the problem~\eqref{eq:opt_bp} is less complex than~\eqref{eq:opt_w} and will instead be solved. Problem~\eqref{eq:opt_bp} can be solved by the branch-reduce-and-bound algorithm whose complexity grows {\em exponentially} with $M+K$~\cite{bjornson13}.

\subsection{Proposed Beamforming Neural Network}
\label{propsedBNN}

We propose to train a second neural network to minimize the negative averaged sum rate as the loss function
\begin{equation}
  \text{Loss} = -\frac{1}{T} \sum_{s=1}^T \sum_{i=1}^{M+K} \log_2(1 + \tilde{\gamma}_i (s))
  \label{eq13}
\end{equation}
where $s$ denotes the training instance and $T$ denotes the total number of training instances.  Unlike the RIS CNN in the previous section, we train this model unsupervised. Therefore, there is no need to create ground truth by solving~\eqref{eq:opt_bp} directly and there is significant reduction in complexity during offline training. 

Table~\ref{tab1} shows the layers of the proposed BNN, which accepts $\bR_{h_l}$, $P_T$, and $\be^{\max}_l$ for $l = 1, 2, \ldots, M+K$ through the input layers 1, 2, and 3, respectively. Complex entries for $\bR_{h_l}$ and $\be^{\max}_l$ are represented by two-dimensional arrays of real numbers.  The inputs are then batch normalized, flattened, and concatenated to feed into the subsequent layers of the BNN.  Batch normalization ensures that the inputs have zero means and unit variances, effectively reducing the internal covariate shift~\cite{IoffeS15}. Flattening, on the other hand, transforms the data into a format suitable for feedforward processing in dense layers. We implemented five dense layers with each of the first three layers followed by a batch normalization layer.  After the three dense layers, we proceed with the first and second lambda layers and the last of the two dense layers. The first half of the third dense layer is the first lambda layer, while the second half is the second lambda layer.  The output of the first and second lambda layers is used to obtain $p_i$ and $\beta_i$, respectively, through the fourth and fifth dense layers.  

To satisfy constraints~\eqref{const_p} and~\eqref{const_b}, the third and fourth lambda layers scale the optimizing variables: 
\begin{equation}
    p^*_i = \frac{P_{T}}{\sum_{l=1}^{M+K} p_l} p_i \ \text{ and } \
    \beta^{*}_i = \frac{P_{T}}{\sum_{l=1}^{M+K} \beta_l} \beta_i. \label{eq15}
\end{equation}
The size of the fourth and fifth dense layers corresponds to the number of optimizing variables $p_i$ and $\beta_i$ where $i = 1,2, \ldots, M+K$. 
We then obtain the beams $\bw_i$ for all $i$ through the fifth lambda layer by applying~\eqref{eq11}. In dense layers, the softplus activation function is applied to ensure positivity constraints~\eqref{const_pos} and to prevent exact zero output errors in the fifth lambda layer~\cite{haozhan15}. The last lambda layer, which functions as the output layer, computes the minimized loss value from~\eqref{eq_appxsinr} and~\eqref{eq13}. 

Similarly to RIS CNN, the computational complexity of the online stage of the BNN can be measured by the total number of FLOPs of all dense layers. The number of FLOPs for a dense layer with $N_I$ input dimensions and $N_O$ output dimensions grows at the rate of $\mathcal{O}(N_I N_O)$. For a system with the number of BS antennas $N_t = 16$ and the number of UEs $M+K = 10$, From Table~\ref{tab2}, the total number of FLOPs for this BNN model is approximately 2.86 million, as shown in Table~\ref{tab2}. We only account for the dense layers in FLOPs calculation as in~\cite{TianLin}.

\begin{table}
\caption{Output dimensions and activation functions of all layers of the proposed BNN.}
\label{tab1}
\begin{center}
\newcolumntype{C}{>{\centering\arraybackslash}X} 
\begin{tabularx}{0.9\columnwidth}{|c|C|c|} 
\hline
\textbf{Layer number} & \textbf{Output dimension} & \textbf{Activation function} \\ \hline
Input layer 1 & $[M+K, 2, N_t, N_t]$ & $\diagdown$  \\ 
\hline
Input layer 2 & $[1]$ & $\diagdown$ \\ 
\hline
Input layer 3 & $[M+K, 2, N_t, 1]$ & $\diagdown$ \\ 
\hline
Concatenate layer & $[2 N_t (M+K)(N_t + 1) + 1, 1]$ & $\diagdown$ \\ \hline
Dense layer 1 & $[256, 1]$ & softplus \\ \hline
Dense layer 2 & $[128, 1]$ & softplus \\ \hline
Dense layer 3 & $[64, 1]$ & softplus \\ \hline
Lambda layer 1 & $[32, 1]$ & $\diagdown$ \\ \hline
Lambda layer 2 & $[32, 1]$ & $\diagdown$ \\ \hline
Dense layer 4 & \([M+K]\) & softplus \\ \hline
Dense layer 5 & \([M+K]\) & softplus \\ \hline
Lambda layer 3 & \([M+K, 1, 1]\) & $\diagdown$ \\ \hline
Lambda layer 4 & \([M+K, 1, 1]\) & $\diagdown$ \\ \hline
Lambda layer 5 & \([M+K, N_t, 1]\) & $\diagdown$ \\ \hline
Lambda layer 6 & \([1]\) & $\diagdown$ \\ \hline
\end{tabularx}
\end{center}
\end{table}

\begin{table}
\caption{The number of FLOPs for each dense layer for BNN with $N_t = 16, M + K = 10$.}
\label{tab2}
\begin{center}
\newcolumntype{C}{>{\centering\arraybackslash}X} 
\begin{tabularx}{0.9\columnwidth}{|c|C|c|c|} 
\hline
\textbf{Layer number} & \textbf{Input size} & \textbf{Output size} & \textbf{Number of FLOPs} \\ \hline
Dense layer 1 & $5,441$ & $256$ & $2,782, 336$ \\ \hline
Dense layer 2 & $256$ & $128$ & $65, 526$ \\ \hline
Dense layer 3 & $128$ & $64$ & $16,384$ \\ \hline
Dense layer 4 & $32$ & $10$ & $640$ \\ \hline
Dense layer 5 & $10$ & $10$ & $200$ \\ \hline
\end{tabularx}
\end{center}
\end{table}

\subsection{Zeroforcing Beamforming With Water-filling Power Allocation}
\label{sec:zf}

We compare the performance of the proposed BNN with the statistical zeroforcing (ZF) beamforming in which each transmit beamforming vector lies in the null space of channel covariance of all interfering UEs~\cite{sdmaIT13,kimSensor}. Although linear ZF beams are suboptimal, they perform close to optimum for a low number of UEs $M+K$ or a large total transmit power $P_T$. If the degree of freedom of the transmission is sufficiently high or more precisely, 
\begin{equation}
\sum_{l=1}^{M+K} \rank(\bR_{h_l}) < N_t,
\end{equation}
transmit beam $\bw_l$ can completely pre-cancel the interference and the approximate sum rate can be computed by
\begin{equation}
    \sum_{l=1}^{M+K} \log_2\left(1 + p_l \frac{\bw_l^\dag \bR_{h_l} \bw_l}{\|\bw_l\|^2} \right).
\end{equation}
We can further improve the above sum rate by applying water filling power allocation.

The main complexity of statistical ZF lies in the eigendecomposition and singular value decomposition (SVD) of the channel covariances. The complexity of each operation increases approximately as $\mathcal{O}(N_t^3)$. Since finding each beam requires a single eigendecomposition and two SVDs~\cite{kimSensor}, the complexity of finding ZF beams is approximately $\mathcal{O}((M+K)N_t^3)$ for a large $N_t$ or large number of UEs.

\section{Numerical Results}
\label{numerical}

For numerical simulations, we generate channel covariances for the direct-path UEs by
\begin{equation}
  \bR_{h_i} = \frac{N_t}{L_i} \bA_i \bA_i^{\dagger} \label{eq3}
\end{equation}
where $\bA_i$ is an $N_t \times L_i$ matrix whose $l$th column is the transmit steering vector $\ba(\theta_{i,l})$ corresponding to the angle-of-departure (AoD), $\theta_{i,l}$ of the $l$th path and $0 \le \theta_{i,l} \le 2\pi$. Under the assumption of a uniform linear array with half-wavelength spacing between antennas, the transmit steering vector for the $l$th path of UE $i$ is given by~\cite{sayeed07}
\begin{multline}
\ba(\theta_{i,l}) = \frac{1}{\sqrt{N_t}}[ 1 \ \mathrm{e}^{-j2\pi\cos(\theta_{i,l})} \ \mathrm{e}^{-j2\pi2\cos(\theta_{i,l})} 
\dots \ \mathrm{e}^{-j2\pi(N_t -1)\cos(\theta_{i,l})}]^T. \label{eq4}
\end{multline}
For RIS UE $i$, the channel gain between the $n_t$th BS antenna and the $n$th RIS element can be modeled by~\cite{nadeem20}
\begin{equation}
[\bG_i]_{n_t, n} = \mathrm{e}^{j\pi [(n_t - 1) \sin(\xi_i(n)) \sin(\upsilon_i(n)) -(n-1) \sin(\xi_i(n)) \sin(\upsilon_i(n))]} 
\end{equation}
where $\xi_i (n)$ for the $n$th element of RIS panel $i$, $N$ denotes the elevation AoD from the BS and is independent and uniformly distributed (i.i.d.) on $[0, \pi]$ for full-rank $\bG_i$, and  $\upsilon_i (n)$ denotes the azimuth AoD for RIS element $n$ and is i.i.d. on $[0, 2\pi]$. The channel covariance between the RIS and UE $i$, $\bR_{g_i}$, can also be modeled by~\eqref{eq3}.

To train RIS CNN, we apply ADMM to find the optimal $\btheta_k$ for each set of $\bG_k$ and $\bR_{g_k}$ for $k = M+1, M+2, \ldots, M+K$. For each sample, we set a maximum of 1000 iterations with a terminating tolerance of $10^{-5}$ and a regularization parameter of 2000. We created 42,500 samples with a validation split of $0.2$ to train our supervised RIS CNN model. For $N_t = 16$ and $N = 30$, the hidden layers consist of 512, 256, and 128 neurons, respectively, with an initial rate of $10^{-3}$ and a batch size of 32. The objective function for problem~\eqref{eq:opt_RIS} of obtained by the proposed RIS CNN is compared with that of ADMM and random RIS phases in Table~\ref{tab5}. With system parameters $N_t = 16$, $N = 30$ and $M+K = 6$, the result is shown for 2 sets of the number of direct UEs and that of RIS-assist UEs. The proposed RIS CNN performs close to ADMM for both sets and significantly better than random RIS phases.  

\begin{table}
\caption{Values of objective function~\eqref{eq:opt_RIS_max} for one sample set of channel covariances with $N_t = 16$, $N =30$, and $\rank(\bR_{g_k}) = 1, \forall k$.}
\begin{center}
\small
\newcolumntype{C}{>{\centering\arraybackslash}X} 
\begin{tabularx}{0.9\columnwidth}{|c|c|C|C|C|} 
\hline
\multicolumn{2}{|c|}{Number of UEs}& \multicolumn{3}{|c|}{Objective function $\btheta_k^{\dag} (\bR_{g_k}^T \odot [\bG_k^{\dag} \bG_k]) \btheta_k$}\\
\hline
$M$ & $K$ & \textbf{Random} & \textbf{ADMM} & \textbf{Proposed RIS CNN}\\ \hline
$5$ & $1$ & $14.1402$ & $49.0778$ & $48.4836$ \\ \hline
$3$ & $3$ & $17.7822$ & $57.4043$ & $55.9989$ \\ \hline
\end{tabularx}
\label{tab5}
\end{center}
\end{table}

For BNN, we generate 50,000 samples of which 85\% are used to train the BNN with a validation split of 0.3 and the remaining 15\% are for testing. RIS coefficients for the samples are randomized.  The size of the validation set is chosen from the experiment to ensure that the model is generalized well to unseen data. The number of direct UEs $M$ and RIS-assist UEs $K$ is fixed for all samples. During training, early stopping is employed to avoid overfitting and increase performance~\cite{Loughrey2005UsingET}. The hidden layers for BNN model consist of 256, 128, and 64 neurons, respectively with an initial learning rate of $10^{-5}$ and a batch size of 32. The source code for RIS CNN and BNN can be accessed on github\footnote{\href{https://github.com/khinthandarkyaw98/Optimization-of-Transmit-Beamforming-and-RIS-Coefficients-Using-Channel-Covariances-in-MISO-Downlink}{The link will be activated when the article is published}.}.

\begin{figure}
\centering
\includegraphics[width=\myfigwidth]{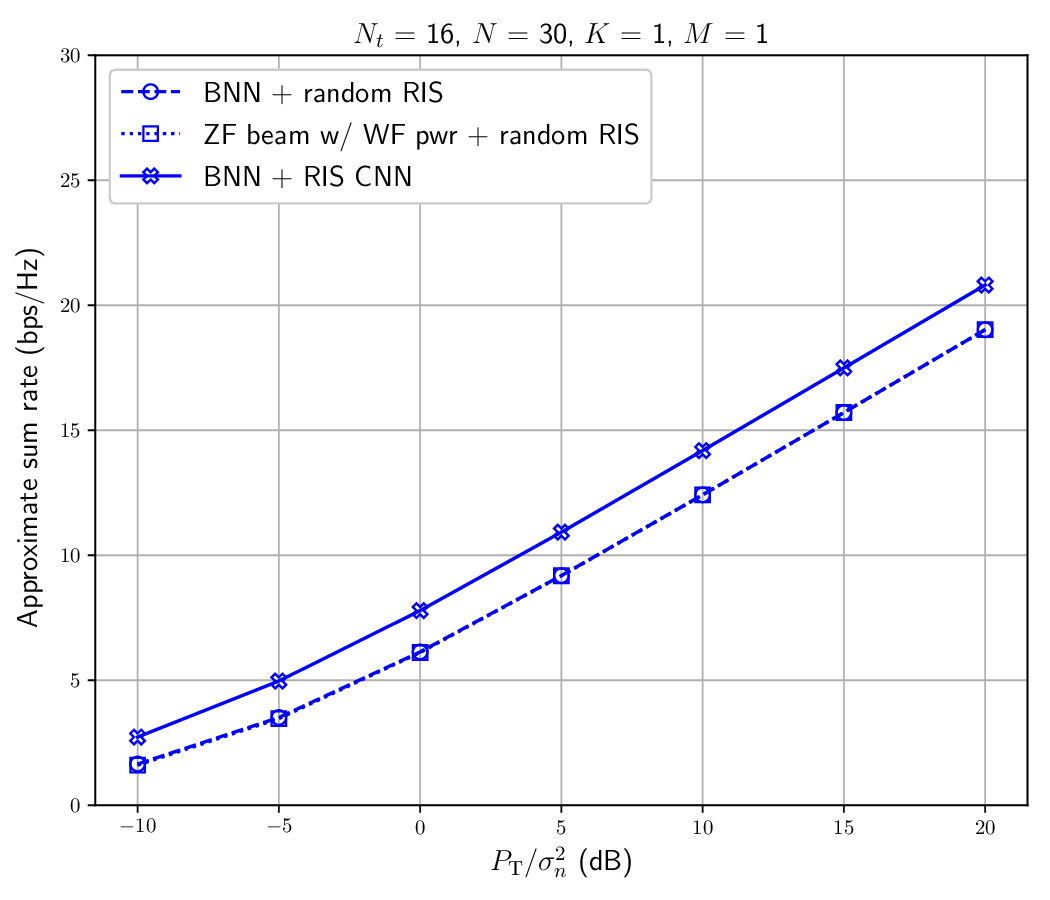}
\caption{Approximate sum rates for BNN and ZF beamforming with random RIS coefficients and BNN with RIS CNN coefficients with $N_t = 16$, $N = 30$, $M=1$, $K=1$, $\rank(\bR_{h_k})=1$, and $\rank(\bR_{g_k})=1, \forall k$.}
\label{fig5}
\end{figure}

To compare the performance of the proposed BNN and ZF beamforming from Sections~\ref{propsedBNN} and~\ref{sec:zf}, respectively, we randomize the RIS coefficients and obtain the resulting sum rate over all UEs in Fig.~\ref{fig5}. For this figure, the number of BS antennas $Nt = 16$, the number of RIS coefficients of an RIS panel $N = 30$, the number of direct UEs $M=1$, and the number of RIS-assist UEs $K = 1$. The figure shows that the proposed BNN and ZF beamforming achieves the same sum rate over different values of SNR. Since ZF beamforming performs close to the optimum when the number of users is small~\cite{marzettaIT13}, we can infer from this setup with only 2 UEs that BNN performs close to the optimal transmit beamforming. In subsequent sets of results with a higher number of UEs, we will see that BNN outperforms ZF beams. If RIS CNN is applied to predict the optimal RIS coefficients, a rate gain of 2 - 3 bits per second per Hz is obtained for all SNR ranges. The percentage rate gain is substantial in a low SNR regime. For the results, we assume sparse channels with rank-one covariance. 

\begin{figure}
\centering
\includegraphics[width=\myfigwidth]{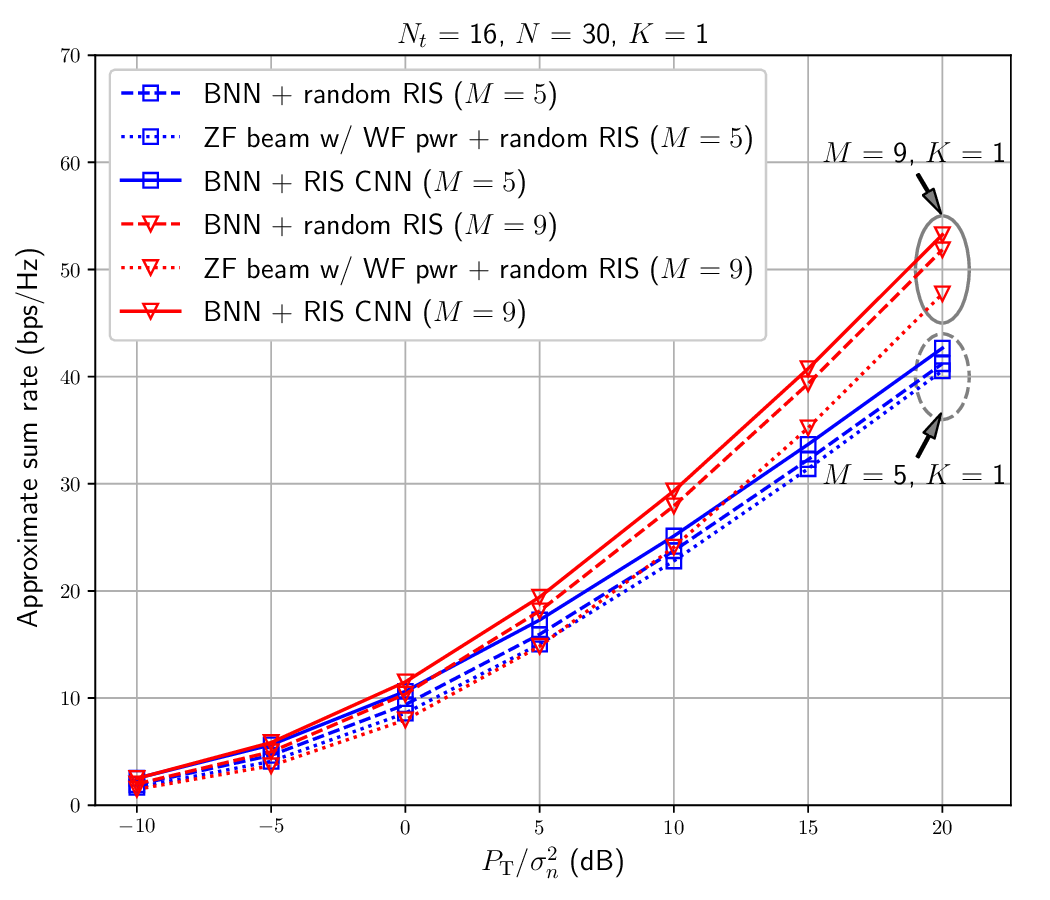}
\caption{Approximate sum rates for BNN with either random RIS or RIS CNN, and ZF beams with random RIS ($N_t = 16$, $N =30$, $K = 1$, $M =5$ or $9$, $\rank(\bR_{h_k})=1$, and $\rank(\bR_{g_k})=1, \forall k$).}
\label{fig2}
\end{figure}

\begin{figure}
\centering
\includegraphics[width=\myfigwidth]{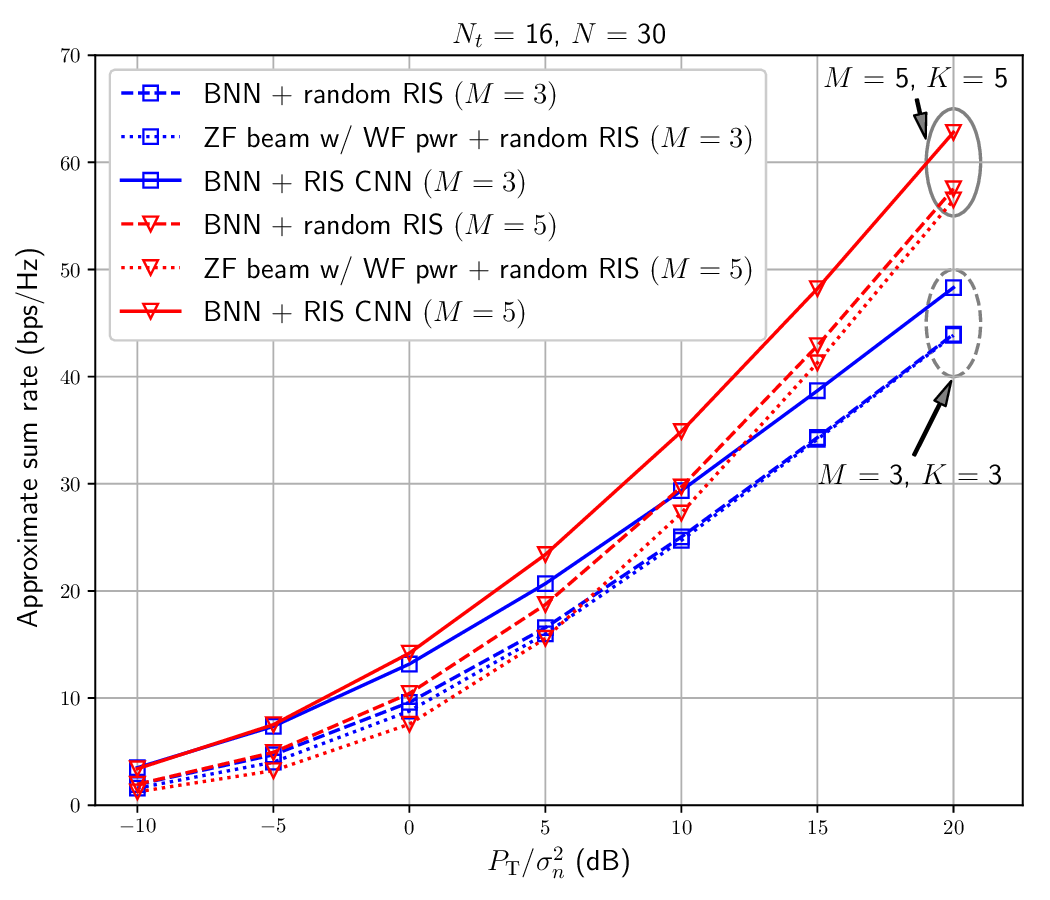}
\caption{Approximate sum rates for BNN with either random RIS or RIS CNN, and ZF beams with random RIS ($N_t = 16$, $N =30$, either $M=5$ and $K = 5$ or $M =3$ or $K=3$, $\rank(\bR_{h_k})=1$, and $\rank(\bR_{g_k})=1, \forall k$).}
\label{fig3}
\end{figure}

For Figs.~\ref{fig2} and~\ref{fig3}, we plot the approximate sum rate versus the total SNR $\frac{P_T}{\sigma^2_n}$ for various user loads, comparing BNN, ZF beamforming, RIS CNN, and random RIS coefficients. For Fig.~\ref{fig2}, there is a single RIS-assist UE while the number of direct UEs is either 5 (blue curves with squares) or 9 (red curves with triangles). Application of both BNN and RIS CNN achieves the highest sum rate for both sets of parameters. However, the sum-rate difference between RIS CNN and random RIS when BNN is applied is small. This can be attributed to one RIS-assist UE in the downlink. In Fig.~\ref{fig3}, the same sum-rate difference is more pronounced since there are more RIS-assist UEs, which contribute more to the overall sum rate. The ZF beams with random RIS are the worst, especially for a large load of UEs or low SNR regimes.

\begin{figure}
\centering
\includegraphics[width=\myfigwidth]{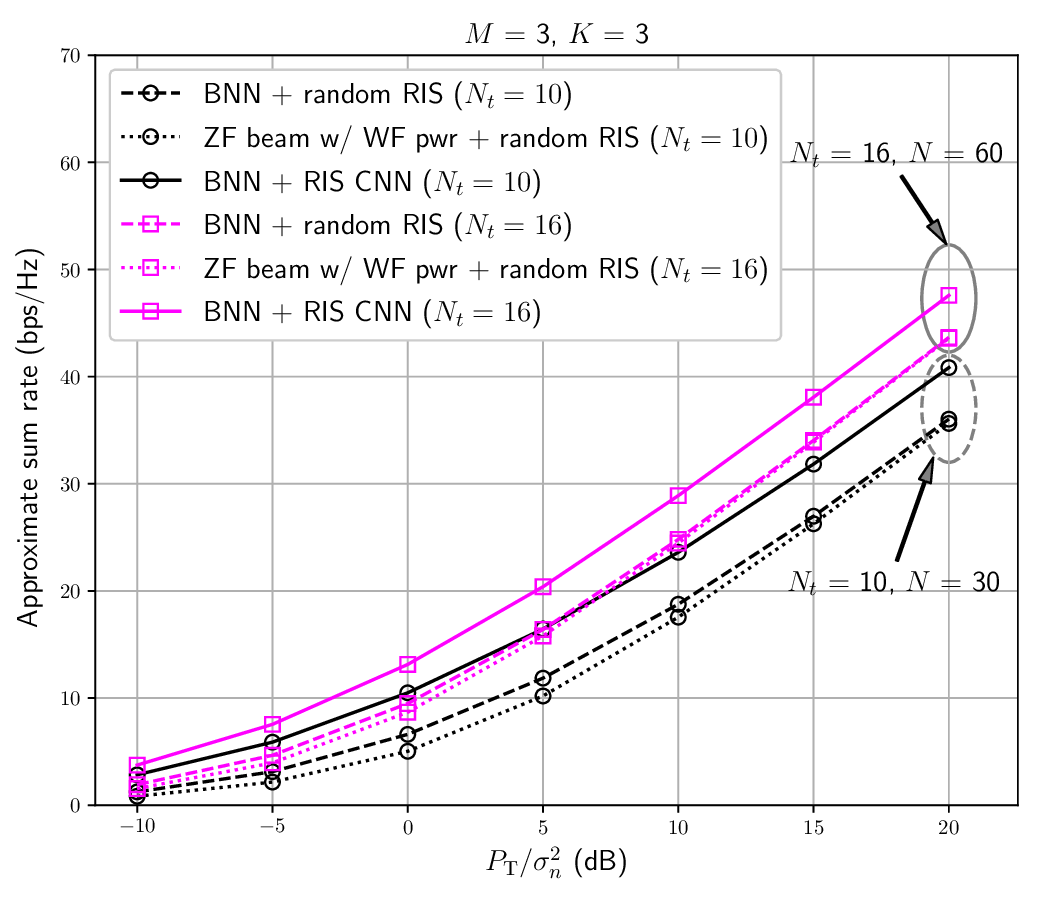}
\caption{Approximate sum rates for BNN with either random RIS or RIS CNN, and ZF beams with random RIS ($M = 3$, $K =3$, either $N_t = 16$ and $N=60$ or $N_t=10$ and $N=30$, $\rank(\bR_{h_k})=1$, and $\rank(\bR_{g_k})=1, \forall k$).}
\label{fig4}
\end{figure}

In Fig.~\ref{fig4}, the number of BS antennas is reduced from 16 to 10. As expected, this reduction worsens the performance of all schemes because there are fewer degrees of freedom to mitigate interference. For a load of $M + K = 6$ UEs and an SNR of $15$ dB, the sum rates of the ZF and BNN beamforming with random RIS coefficients are approximately 5 bits per second per Hz lower than that of BNN with RIS CNN. This highlights the effectiveness of the RIS CNN approach even with fewer BS antennas. Compared to the sum rates in Fig.~\ref{fig3}, we see a slight rate decrease when the number of reflecting elements for each RIS panel increases from $N = 30$ to $60$. Since there are more variables to predict, RIS CNN returns slightly less accurate coefficients.

\begin{figure}
\centering
\includegraphics[width=\myfigwidth]{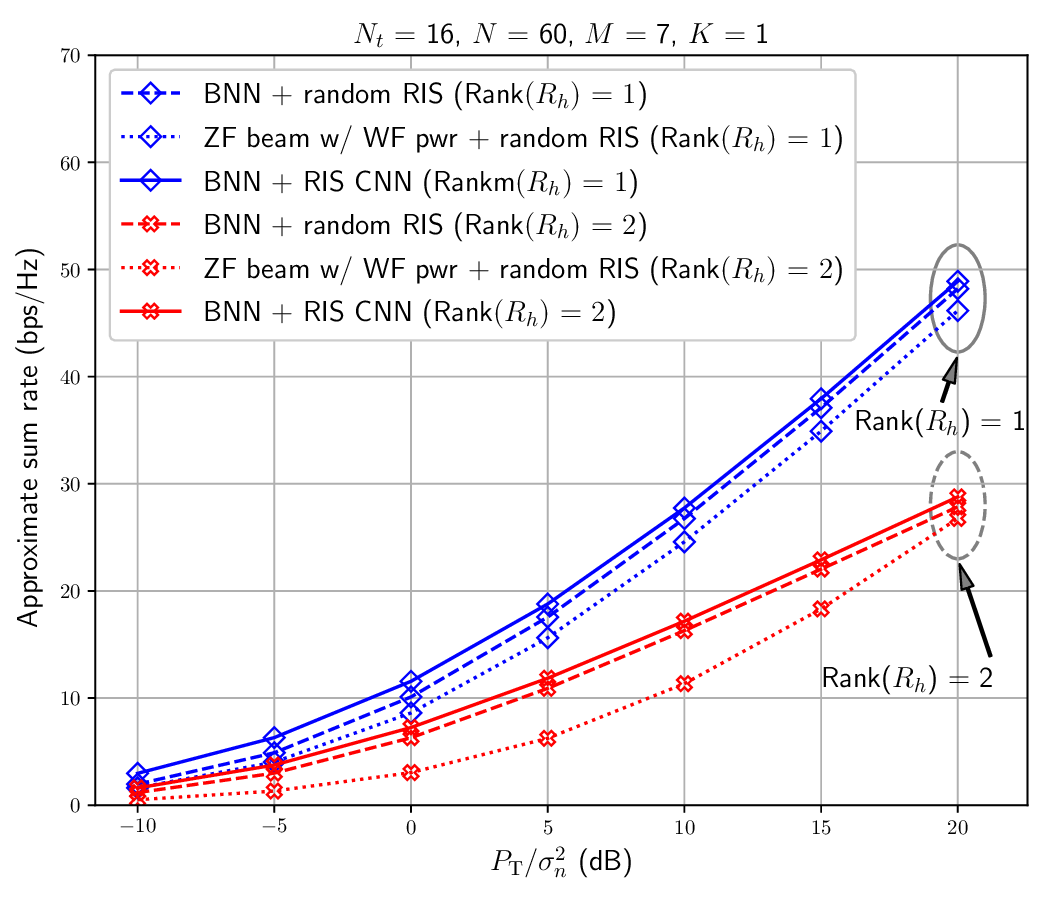}
\caption{Approximate sum rates for BNN with either random RIS or RIS CNN, and ZF beams with random RIS ($N_t = 16$, $N =60$, $M=7$, $K=1$, $\rank(\bR_{g_k})=1$, and either $\rank(\bR_{h_k})=1$ or $\rank(\bR_{h_k})=2, \forall k$).}
\label{fig6}
\end{figure}

For Fig.~\ref{fig6}, we set $N_t=16$, $N=60$, $M=7$, and $K=1$ with the rank of $R_{h_k}$ and $R_{g_k}$ being 1 or 2 for all UEs. With increased rank, performance suffers because there are fewer degrees of freedom in signal space to avoid interference. However, ZF beamforming suffers the largest rate loss especially for moderate or low SNR when compared to BNN and BNN with RIS CNN. We remark that the earlier version of this figure was presented at ECTI-CON 2024 with an error of curve labeling.  The error has been corrected in Fig.~\ref{fig6}.

\begin{figure}
\centering
\includegraphics[width=5in]{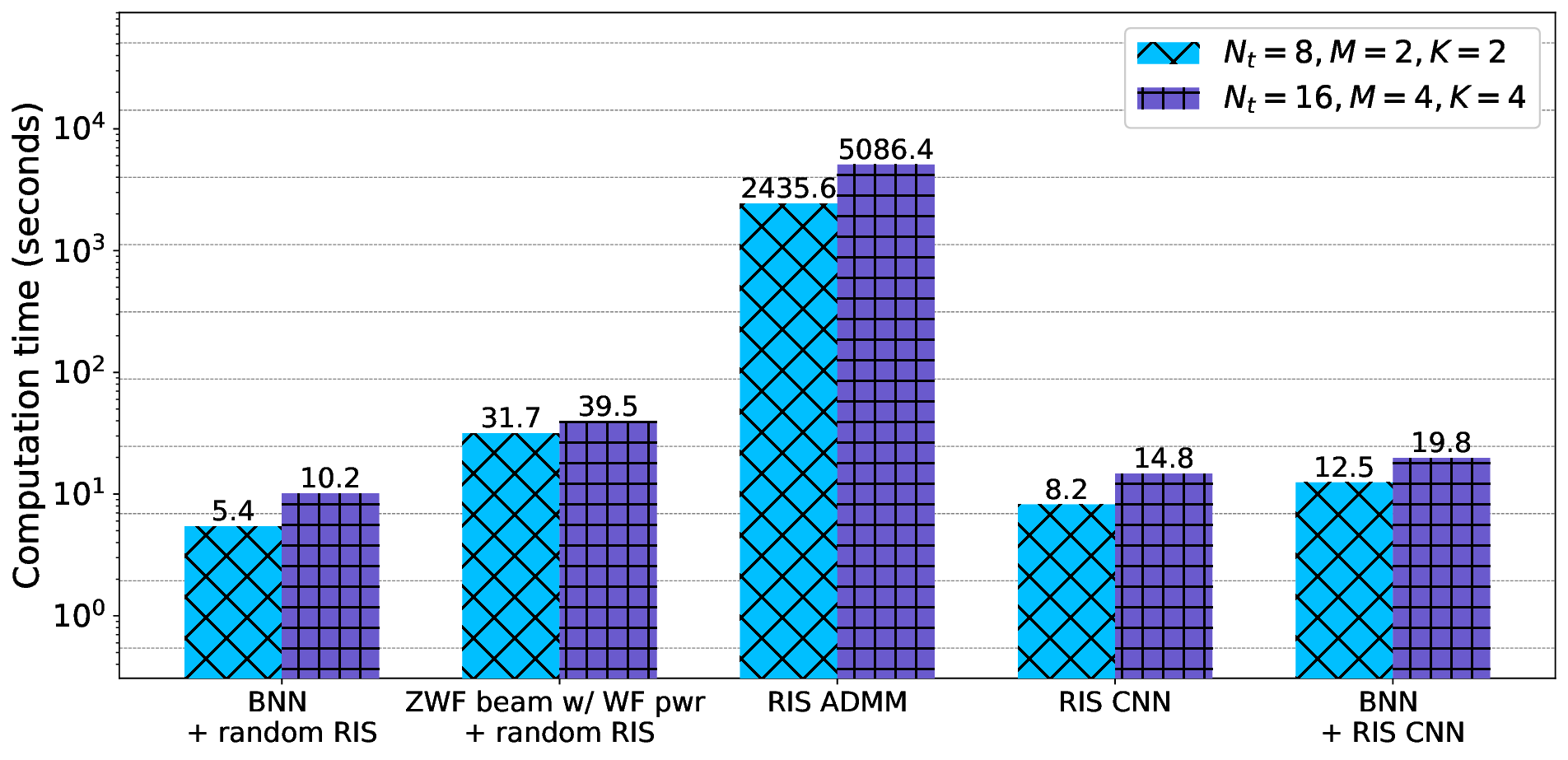}
\caption{The computation time in seconds for 7000 channel samples is shown for all schemes ($N = 30$, $\rank(\bR_{h_k})=1$, and  $\rank(\bR_{g_k})=1, \forall k$, either $N_t = 8$ and $M+K = 4$ or $N_t = 16$ and $M+K = 8$).}
\label{fig_bar}
\end{figure}

To compare computational complexity for different schemes,  Fig.~\ref{fig_bar} shows the total computation time for 7,000 channel instances using the same hardware\footnote{MacBook Air with M1 chip (8-core CPU and 7-core GPU) and 8 GB RAM.}. Since BNN and RIS CNN training are carried out offline, we do not account for the time spent training those models. We consider 2 systems with $N_t = 8$ or $16$ and half UE load $(M+K)/N_t = 0.5$. For RIS coefficients, the time for ADMM is 2 orders of magnitude larger than that for RIS CNN. For BS beams, the ZF scheme takes many times longer than the proposed BNN. The combination of BNN and RIS CNN takes a bit more time than either BNN or RIS CNN, but takes less time than ZF scheme and significantly less than ADMM. As anticipated, the computation time increases across all schemes when the system size is doubled.

\section{Conclusions}

Our proposed BNN and RIS CNN achieves a higher sum rate than statistical ZF beamforming for heavy load, especially for a larger number of RIS-assist UEs. The computational complexity of the proposed scheme is less than that of the ZF scheme and significantly lower than that of ADMM. The proposed schemes utilize only channel covariances of UEs, which do not change often. Hence, the transmit beams and RIS coefficients do not need frequent updates. However, if the channel information used during the training of the BNN or RIS CNN models is not accurate, the sum rate performance will suffer. Another direction to explore is to deploy active RIS panels instead of passive ones, as assumed in this work. Active RIS has recently been shown to significantly increase the performance of wireless channels.

\section{Acknowledgment}

A preliminary version of this work has been presented at ECTI-CON 2024, in Khon Kaen, Thailand~\cite{ecti24}. This work is funded by Thailand Advanced Institute of Science and Technology (TAIST), National Science and Technology Development Agency (NSTDA), Faculty of Engineering, Kasetsart University (KU), and Tokyo Institute of Technology under the TAIST Tokyo Tech Program and by Thailand Science Research and Innovation Fund, Chulalongkorn University under Grant CU\_FRB65\_ind (12)\_160\_21\_26.

\bibliographystyle{elsarticle-num} 
\bibliography{IEEEfull,ris}

\end{document}